\documentstyle[epsfig]{aipproc}

\begin{document} 

%\noindent To be published in {\it Young Supernova Remnants:
%Eleventh Astrophysics Conference, College Park, Maryland, October
%2000} (AIP Conference Proceedings), eds. S. S. Holt and
%U. Hwang (American Institute of Physics, Melville, NY), as part of the
%proceedings of the {\it Eleventh Annual October Astrophysics
%Conference in Maryland -- Young Supernova Remnants},
%College Park, Maryland, October 16-18, 2000.

\title{High Resolution Spectroscopy of Two Oxygen-Rich SNRs with the Chandra HETG}

\author{C.R.~Canizares, K.A.~Flanagan,  D.S.~Davis, D.~Dewey, J.C.~Houck,
M.L.~Schattenburg}
\address{Center for Space Research, Massachusetts Institute of Technology\\
Cambridge, MA 02139}

%\lefthead{LEFT head}
%\righthead{RIGHT head}
\maketitle
\begin{abstract}
The HETG can be used to 
obtain spatially resolved spectra of moderately extended sources.
We present preliminary results for two well studied,
oxygen rich supernova
remnants in the Magellanic clouds, E0102-72 and N132D.  The dispersed
spectrum of E0102-72 shows images of the remnant in the light of individual
emission lines from H-like and He-like ions of O, Mg, Ne and He-like Si
 with no evidence
of Fe. The diameters of the images for various ions,
 measured in the cross-dispersion
direction, increase monotonically with the ionization age for
the given ion. This shows in detail the progression of the reverse shock
through the expanding stellar ejecta. We see clear evidence for asymmetric
Doppler shifts across E0102-72 of $\sim \pm2000~km~s^{-1}$. These can be
modelled approximately by a partially-filled, expanding shell inclined
to the line of sight.  The dispersed spectrum of N132D is more affected
by spatial/spectral overlap but
also shows monochromatic images in several strong lines. Preliminary spectra
have been extracted for several bright knots. Some regions of
oxygen-rich material, presumably stellar ejecta, 
are clearly identified. Additional details on E0102-72 are presented
by Flanagan {\it et al.}~and Davis {\it et al.}~in these proceedings,
and further analysis is in progress.
\end{abstract}

\section*{Introduction}

The High Energy Transmission Grating (HETG) consists of an array of
periodic nanostructures that can be inserted behind the {\it Chandra}
mirrors in
order to disperse the focused X-ray beam into a spectrum at the focal
plane.  The HETG array includes two grating types, MEG and HEG, which
together cover the energy band 0.4-8 keV with resolving powers of up to
~1000.  The dispersed spectrum is read out by the Advanced CCD Imaging
Spectrometer (ACIS-S). A complete description is given in \cite{Canizares01}
and details can be found in the {\it Chandra} Proposers guide and at
{\tt http://space.mit.edu/HETG}.

As a
``slitless" dispersive spectrometer, the HETG is most straightforward
when used to observe point sources: a zeroth order image is formed at the
normal focal point, with dispersed spectra on either side (the HEG and MEG
spectra are offset from one another by 10 degrees to form a shallow "X"
centered on zeroth order). For point sources, the resolution of the
complete spectrometer is 0.022~\AA~ and 0.012~\AA~ for MEG and
HEG, respectively.  For moderately extended sources, the spatial and
spectral information are mixed in the dispersion direction, but not in the
cross dispersion direction.  Although this mixing complicates the analysis,
moderately extended sources like those presented here are very amenable to
simultaneous spectral/spatial analysis, particularly if their emission is
dominated by distinct spectral lines.  In that case the dispersed spectrum
is analogous to a spectro-heliogram, showing a series of monochromatic
images of the source in the light of individual spectral lines.

This paper gives a brief report of HETG observations of two relatively
compact supernova remnants (SNRs) in the Magellanic clouds, E0102-72 and
N132D. Both are members of the oxygen-rich class of SNRs, of which Cas A is
often taken as the prototype and which are thought to be products of Type
1b or II supernovae in massive stars \cite{Blair00}.
More detailed reports of our results on E0102-72 
are included in the proceedings of this meeting
(see \cite{Flanagan01} \cite{Davis01}).

\section*{E0102-72}

The SNR 1E0102.2-7219 is a well studied member of the oxygen rich class of
supernova remnants located in the SMC. It has a radius of ~20 arc sec
(6.4 pc), which is a perfect size for spectral imaging with the HETG.
Recently, moderate resolution
X-ray spectra integrated over the remnant were obtained with ASCA 
\cite{Hayashi94}.  Gaetz {\it et al.} \cite{Gaetz00} reported spectrally 
resolved imaging from {\it Chandra's} 
ACIS detector, which shows an almost classic, text-book SNR with
a hotter outer ring identified with the forward shock surrounding a cooler,
denser inner ring which is presumably the reverse-shocked stellar ejecta.
Hughes {\it et al.} \cite{Hughes00} combined 
the {\it Chandra} image with earlier {\it Einstein} and
{\it ROSAT} images to measure X-ray proper motions, which give an expansion age
of 1000 yr, consistent with earlier estimates (they also deduce that
a significant fraction of the shock energy has gone into cosmic rays).
  Blair {\it et al.} \cite{Blair00}
analyzed extensive observations with HST of the optical/UV filaments. At this
meeting, Eriksen {\it et al.} \cite{Eriksen01} report the analysis of new
Fabry-Perot observations that yield an age of $\sim$2100 yr for the
remnant.

We observed E0102 for a total of ~140 ksec on two occasions with the
{\it Chandra} HETG.  
A portion of the dispersed spectrum is shown in Fig. 1.  It
shows multiple images of the SNR in the light of individual spectral lines
from various ions.  Most prominent are the lines of H-like and He-like O,
Ne, Mg.  We see no clear evidence for any Fe emission (although weak Fe
lines are reported by Rasmussen {\it et al.} \cite{Rasmussen01} 
at this meeting from an observation
with the {\it XMM-Newton} Reflection Grating Spectrometer).  The O {\sc viii} image alone
suggests that this remnant contains at least several solar masses of oxygen
in the reverse-shocked ejecta.

\begin{figure*}[h!] %
\centerline{\epsfig{file=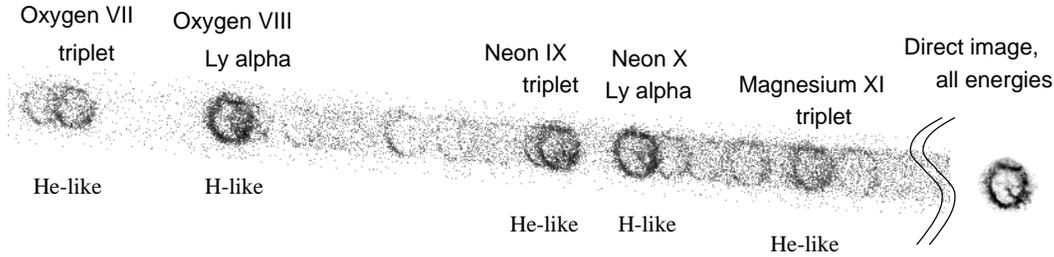,angle=-90,width=5.5in}}
\vspace{10pt}
\caption{Dispersed high resolution X-ray spectrum of E0102-72.
Shown here is a portion of the -1 order formed by the medium energy gratings
(MEG).
The zeroth order, which combines all energies in an undispersed image is 
at the right. Images formed in the light of several strong X-ray emission
lines are labeled.}
\label{fig1}
\end{figure*}

\subsection*{Imaging the Reverse Shock}

The time scale for the ions in E0102 to reach ionization equilibrium is
comparable to or longer than its age, so one would expect the plasma not
to have achieved ionization equilbrium \cite{Hughes85}. Hayashi {\it et al.} \cite{Hayashi94}
found evidence
for departures from ionization equilibrium for the integrated spectrum of
the SNR (which ASCA could not resolve spatially) based on the relative
intensities of He-like and H-like lines.  They could not find a consistent
model to explain the line ratios for different ions, even allowing for
multiple components, and concluded that abundance inhomogeneities must be
present. If we integrate the flux over the remnant for several of the 
strong, relatively isolated lines, we also find line ratios that indicate
departures from ionization equilibrium \cite{Davis01}.

Gaetz {\it et al.} \cite{Gaetz00} find direct evidence for progressive
ionization of oxygen in the spectrally resolved {\it Chandra} image. They
compared monochromatic images (at CCD spectral resolution) at the energies
of the O {\sc vii} and O {\sc viii} lines, and found that the O {\sc vii} 
emitting region lies
inside that of O {\sc viii}.  This is what one would expect if the ejecta are
being subject to a reverse shock propagating backwards (in a Lagrangian
sense) towards the center of the remnant (ionization at smaller radii lags
behind the ionization at larger radii).

We observe this progressive ionization quantitatively by comparing
the images from the He-like resonance lines and H-like Ly$\alpha$ lines of
O, Ne and Mg, and of He-like Si. Details are given elsewhere in these 
proceedings \cite{Flanagan01}, but the qualitative evidence can be seen
directly from Fig 1. For both O and Ne, the images of the He-like lines are
noticeably smaller than the H-like images. 
Cuts across the spectrum in the
cross-dispersion direction (to avoid any spectral/spatial confusion) show
that, for each ion, the diameter of the He-like image is smaller than that
of the H-like image.  Furthermore, for a given temperature, one can find a
characteristic ionization time scale or ionization age 
(ionization age is given by 
$\tau~=~n_et$ for electron density $n_e$ and time $t$) at
which a given ion reaches its peak population fraction, and therefore its
peak emissivity in an ionizing plasma.  Doing this for the ions in question
(assuming logT=7.05) shows a monotonic increase in the diameter of the
emitting region with increasing characteristic ionization timescale
This suggests a simple interpretation, in which the differences
in image diameters for the different ions are caused entirely by the
progressive passage of the reverse shock through relatively homogeneous
stellar ejecta.  A more comprehensive analysis, which will take into
account the likely density gradient in the ejecta, for example, is now
underway.

\subsection*{Differential Doppler Shifts}

We have found clear evidence for Doppler shifts in several of the emission
lines, and evidence that these vary systematically across the remnant.
This can be seen from a comparison of the two (plus and minus order)
dispersed images from a strong line, such as Ne~{\sc x}~Ly~$\alpha$, with
each other and with the
zeroth order image (constructed by selecting a narrow range of
 ACIS pulse heights at the corresponding line energy). For example,
even in Fig. 1 it is clear that the Ne {\sc x} image appears slightly elongated
in the dispersion direction compared to the zeroth order image.

The fact that the HETG has both plus and minus order images breaks the
spectral/spatial degeneracy that would otherwise confuse the signature
of a Doppler shift.
A dispersed image in the light of a single line that is distorted due to
intrinsic spatial variations will look identical on either side of zeroth
order, whereas a distortion due to a wavelength (Doppler) shift will appear
with opposite offsets in the plus and minus orders, (i.e. a shift to longer
wavelength moves to the right in the plus order but to the left in the
minus order image, showing reflectional symmetry about zeroth order).
Furthermore, comparison of plus and minus orders also allows one to
identify and avoid confusion from the overlapping images of nearby lines.

We have examined the relatively clean dispersed $\pm1^{st}$ order images 
in the Ne {\sc x}
Ly$\alpha$ line with the corresponding zeroth order image
 and find clear evidence for distortions
with the mirror symmetry expected for Doppler shifts in
wavelength (some positive, others negative).  More than a dozen bright regions
around the remnant were analyzed to determine the sense and magnitude of
the shifts.

The result of the Doppler analysis for Ne {\sc x} Ly $\alpha$ is shown in Fig. 2,
where the arrows pointing left indicate blue-shifts and those pointing
right indicate red-shifts.  The magnitudes of the shifts range from -1600
to +2300 $km~s^{-1}$. The velocity structure in the X-ray is systematic and
clearly asymmetric, showing only redshifts on the eastern side but both red
and blue shifts on the western side. The velocities measured for some of
the optical knots, which generally lie interior to the brightest portions
of the X-ray bright ejecta, are comparable to those found here and also
show complex, asymmetric structure \cite{Tuohy83}

\begin{figure}[h!]
%\begin{center}
%\begin{minipage}[h]{2.5in}
\centerline{\epsfig{file=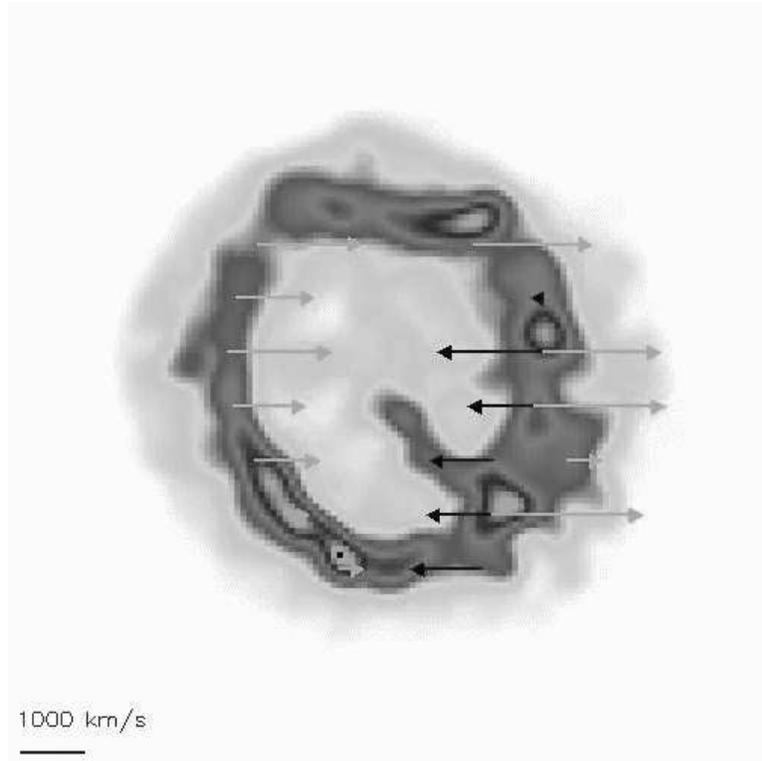,width=4.0in}}
%\plotone{pic_vel_bw.ps}
%\end{minipage}
\vspace{10pt}
\caption{Doppler velocities of  E0102-72 Ne {\sc x} Ly~$\alpha$. 
The lengths and locations of the arrows indicate 
the relative velocities and approximate locations of red/blue shifted
ejecta. Arrows pointing left represent blue shifts;
arrows pointing right represent red shifts.}

%\end{center}
\end{figure}

We have made a preliminary effort at modeling the asymmetric velocity
structure shown in Fig 2.  Rough correspondence is achieved if the emission
comes from a partially filled spherical shell with azimuthal symmetry whose
axis is inclined to the line-of-sight.  The inner radius of the shell is
4.3 pc, its thickness is 1.1 pc, and the velocity increases with radius.
The emissivity is concentrated toward the
equatorial plane (we assume a Gaussian distribution in the cosine of the
latitude on the spherical shell with $\sigma$=0.35).  Interestingly, Hughes
\cite{Hughes88} deduced that the surface brightness distribution measured
with the {\it ROSAT} HRI also indicated that the
emission was concentrated in a thick ring rather than a spherical shell.

\section*{N132D}

N132D is located in the LMC at a distance of 50 kpc, with a radius of $\sim$50
arc sec ($\sim$12 pc) and an estimated age of 3000 yr.  Several recent X-ray studies
have been presented from Beppo-SAX \cite{Favata97} and ASCA \cite{Hughes98}.
Previously, Hwang {\it et al.}\cite{Hwang93} 
performed a non-equilibrium ionization analysis using
moderate resolution spectra from the Solid State Spectrometer combined with
high resolution spectroscopy of selected lines obtained with the Focal
Plane Crystal spectrometer (both instruments were on the
{\it Einstein} observatory).  All these studies suggest the overall
elemental abundances for N132D are roughly similar to those for the LMC as
a whole.  This indicates that N132D is dominated by swept-up material
rather than ejecta, and is thus at a later stage of evolution than
E0102-72. The most recent optical/UV data  \cite{Blair00} support
this conclusion.

\begin{figure}[b!] % fig 3
%\begin{center}
\centerline{\epsfig{file=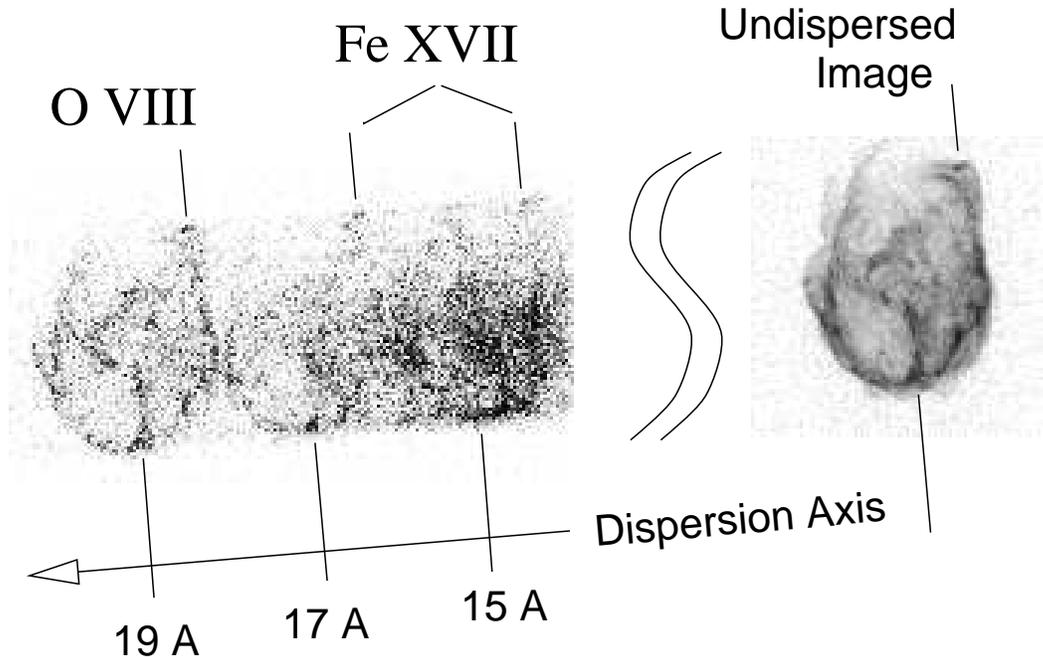,width=5.5in}}
\vspace{10pt}
%\begin{figure}
%\plotone{n132d_Fe17_O8.eps}
\caption{Dispersed high resolution X-ray spectrum of N132D.
A portion of the MEG spectrum is shown at left, with zeroth order
image at right.  }
\label{fig3}
%\end{center}
\end{figure}

We observed N132D for 100 ksec with the {\it Chandra} HETG 
with a roll angle that
places the dispersion direction parallel to the narrow axis of this
U-shaped SNR.  A portion of the MEG spectrum is shown in Fig 3.  The
combination of larger size and more numerous emission lines in N132D result
in greater spectral/spatial overlap than in E0102-72.  Nevertheless, we can
distinguish clear monochromatic images in the light of individual emission
lines of oxygen, neon and iron, for example (iron is largely responsible
for the enhanced number of lines). Even a preliminary examination indicates
strong spatial variations in the concentrations of these species, with some
features richer in iron and others in oxygen. The fact that we see
oxygen rich material is particularly significant, since, contrary
to the situation in E0102, no direct
X-ray evidence for oxygen
rich ejecta had previously been seen \cite{Hughes98} (Behar {\it et al.}
\cite{Behar01} also
see regions of enhanced oxygen with {\it XMM-Newton})

\begin{figure*} % fig 4
%\begin{center}
\centerline{\epsfig{file=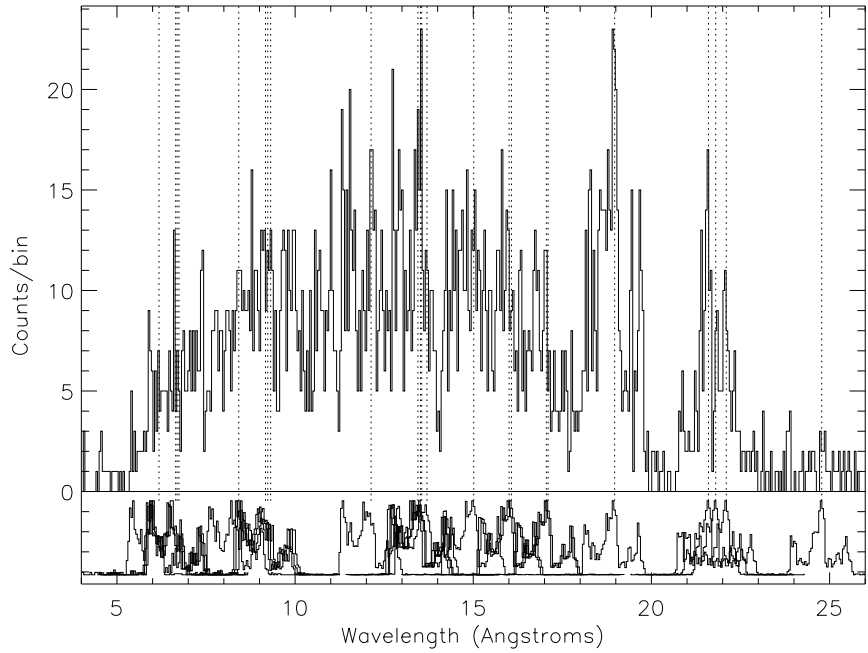,angle=90.,width=5.0in}}
\vspace{10pt}
%\begin{center}
\centerline{\epsfig{file=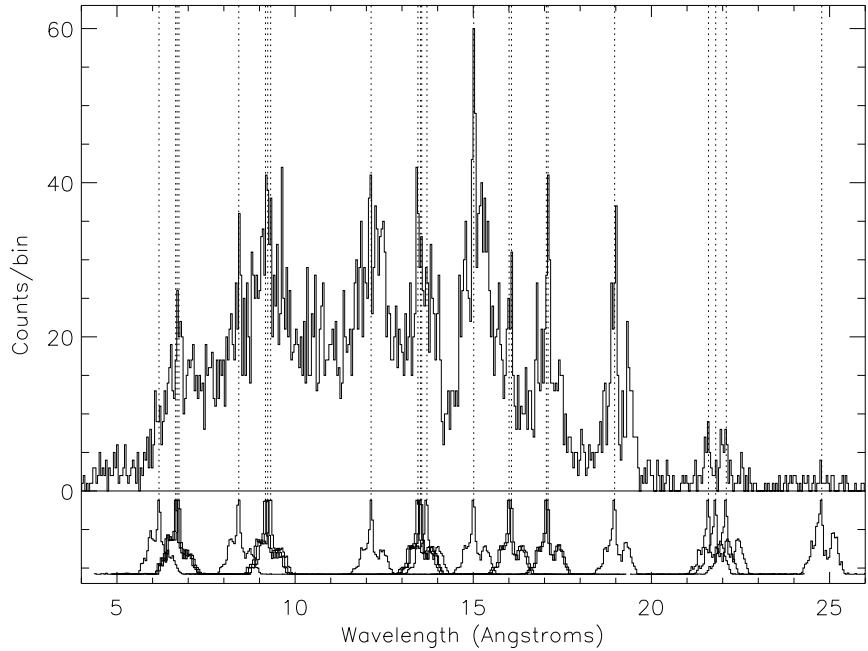,angle=90.,width=5.0in}}
\vspace{10pt}
\caption{``Convolved'' spectra (see text for explanation) of selected
X-ray bright features in N132D Top: a knot in the uppermost part of the
complex of filaments in the interior (it is the topmost filament that 
appears prominently above the lower vertical 
marker at $\sim$19~\AA~ in Fig 3. Bottom:
a knot in the bottom outer edge (marked with the lower vertical 
line in the zeroth order image 
and in several dispersed images)  
that appears enhanced at $\sim$17~\AA~ in Fig 3.
Spectra are uncorrected for instrumental
efficiency. Under each spectrum the effective line shape function
is plotted centered at the
wavelengths of selected emission lines (marked with vertical dotted lines)
which are likely to be strong.}
\label{fig4}
%\end{center}
\end{figure*}

We have begun to measure the spectra of selected knots in N312D by extracting
slices through the dispersed spectral 
image centered on the specific features of
interest. Typically, a given feature will have some ``tilt'' with respect
to the dispersion direction, so spectra in narrow slices perpedicular to
the dispersion direction are shifted to a common origin (e.g. a stack of
slices through the dispersed images of a curved feature whose zeroth
order is shaped like `('
is shifted slice-by-slice and aligned so that its zeroth order becomes
 a straight line perpendicular to
the dispersion direction). The stacked, shifted slices of dispersed spectra 
are then projected onto the dispersion axis (so the zeroth order of our example
would be projected to a point). 
Spatial extent in the dispersion direction within
each narrow slice (e.g., if the `(' varies in thickness) is
not removed, so the projection of the shifted blob onto the dispersion
axis becomes the equivalent line shape function. The resulting spectrum is thus
the true spectrum convolved with this equivalent line shape function (if
there are strong gradients of abundance or temperature across the knot, then
of course the line shape function could be different for different spectral
lines). The next step, not yet undertaken,
is to deconvolve this shape function from the spectrum.

We display in Fig. 4 such ``convolved'' MEG spectra 
extracted for two different X-ray bright features in N132D one from the
interior (top panel) and the other from the rim (bottom panel)
 of the SNR image (see figure
caption). Below each
spectrum, the effective line shape function is plotted repeatedly 
centered at the
wavelengths of selected emission lines that are likely to be strong in N132D. 
The differences in composition of the two blobs are evident.
The interior knot shows strong O~{\sc viii}~Ly~$\alpha$ emission 
at $\sim$19 \AA~ and a prominent O~{\sc vii} triplet at $\sim$22~\AA, whereas
the rim feature shows relatively weak O emission and a strong
Fe {\sc xvii} 2p-3d lines at 15 and 17~\AA.  We
anticipate being able to do a very detailed analysis of the physical
conditions and  composition of many such features.

\section*{Conclusions}

The remnants Cas A, E0102-72 and N132D, with respective ages of ~300~yr,
~1000-2000~yr, and ~3000~yr, provide an interesting sequence of SNRs at
progressive stages in their evolution.  While they have clear differences,
such as the presence of O-burning products in the ejecta of Cas A but not
of the other two (see the discussion in \cite{Blair00}), they also share
significant similarities.  All show oxygen rich optical filaments and, with
the evidence described here for N132D, all show regions of enhanced oxygen
X-ray emission.  E0102-72 provides the first
clear evidence for a progressive reverse shock that is also implied in the
other two SNRs from non-equilibrium analyses. 

 The two younger members, Cas A and E0102-72, show
asymmetric Doppler velocities in the X-ray. Markert
{\it et al.} \cite{Markert83} discovered the Cas A Doppler velocities and asymmetry using
the
{\it Einstein} Focal Plane Crystal spectrometer, and these were subsequently
confirmed and mapped with ASCA \cite{Holt94}, with moderate
angular resolution. The full velocity range for Cas A is roughly half that seen
in E0102, which could well be the result of differences in projection.
Thus, the only two young,
oxygen-rich SNRs for which such measurements could have been made show the
same kind of kinematic behavior.  Our approximate model for E0102, like that
of \cite{Markert83} for Cas A, suggests that the bright ejecta occupy
a partially filled spherical shell in a ring-like geometry (as also suggested
by Hughes \cite{Hughes88} for E0102). While it is possible that the ejecta themselves were expelled
in a ring, it seems more plausible to attribute the geometry to a ring-like
distribution of circumstellar material which is decelerating and shocking
a nearly spherical shell of ejecta, causing it to appear non-spherical. In
that case, a significant fraction of the ejecta would still be relatively
faint in X-rays. This would fit with the suggestion of 
Blair {\it et al.} \cite{Blair00}, who argue that E0102 and N132D may be the 
remnants of very massive O stars that underwent substantial mass loss of their
outer layers prior to exploding as Type Ib supernovae.

 Further study study of these
three bright objects is likely to give us new insights into the detailed
structure and evolution of the remnants of supernovae in massive stars.

\section*{Acknowledgements}
We thank the other members of our HETG/CXC group at MIT for their
many contributions. This work was 
supported by NASA contract NAS8-38249 and SAO SV1-61010.


\begin{references}

\bibitem{Canizares01}Canizares, C.R., {\it et al.}~{\it in preparation}.

\bibitem{Blair00}Blair,~W.P., Morse,~J.A., Raymond,~J.C., Kirshner,~R.P., 
Hughes,~J.P., Dopita,~M.A., Sutherland,~R.S., Long,~K.S. \& Winkler,~P.F., 
{\it ApJ}, {\bf 537}, 667 (2000).

\bibitem{Flanagan01}Flanagan, K.A., Canizares, C.R., Davis, D.S., Dewey, D.
Houck, J.C.,
Schattenburg, M.L., {\it these proceedings}.

\bibitem{Davis01}Davis, D.S., Flanagan, K.A., Houck, J.C., 
Allen, G.E., Schulz, N.S., Dewey, D., Schattenburg, \& M.L., 
{\it these proceedings}.

\bibitem{Hayashi94}Hayashi,~I., Koyama,~K., Masanobu,~O., Miyata,~E., 
Tsunemi,~H., Hughes,~J.P. \& Petre,~R., {\it PASJ}, {\bf 46}, L121 (1994).

\bibitem{Gaetz00}Gaetz,~T.J., Butt,~Y.M., Edgar, R.J., Eriksen,~K.A., 
Plucinsky,~P.P., Schlegel,~E.M. \& Smith,~R.K., {\it ApJ}, {\bf 534}, L47 (2000).

\bibitem{Hughes00}Hughes,~J.P., Rakowski,~C.E. \& Decourchelle, A. 
{\it ApJ}, (2000) in press.

\bibitem{Eriksen01}Eriksen, K.A. {\it et al., these proceedings}.

\bibitem{Rasmussen01}Rasmussen,~A. {\it et al.}, {\it  these proceedings}.

\bibitem{Hughes85}Hughes, J.P. \& Helfand, D.J., {\it ApJ}, 
{\bf 291}, 544 (1985).

\bibitem{Tuohy83}Tuohy, I.R. \& Dopita, M.A.,{\it ApJ},{\bf 268}, L11 (1983)

\bibitem{Holt94}Holt, S.S., Gotthelf, E.V., Tsunemi, H. \& Negoro, H.,
{\it PASJ}, {\bf 46}, L151 (1994).

\bibitem{Hughes88}Hughes,~J.P. 1988, in {\it Supernova Remnants and the 
Interstellar Medium}, ed. R.S.~Roger \& T.L.~Landecker 
Cambridge: Cambridge Univ. Press, p.~125.

\bibitem{Favata97} Favata, F., vink, J., Parmar, A.N., Kaastra, J.S., \& 
Mineo, T., {\it A\&A}, {\bf 324}, L45, (1997).

\bibitem{Hughes98}Hughes, J.P., Helfand, D.J.Hayashi, I., \& Koyama, K.,
 {\it ApJ}, 
{\bf 505}, 732 (1998).

\bibitem{Hwang93}Hwang, U., Hughes, J.P., Canizares, C.R. \& Markert, T.H., {\it ApJ}, {\bf 414}, 219, (1993).

\bibitem{Behar01}Behar, E.  {\it these proceedings}.

\bibitem{Markert83}Markert, T.H., Canizares, C.R., Clark, G.W., Winkler, P.F.,
{\it ApJ}, {\bf 268}, 134 (1983).


\end{references}
\end{document}